\documentclass[conference]{IEEEtran}
\IEEEoverridecommandlockouts

\usepackage[utf8]{inputenc}
\usepackage[T1]{fontenc}

\usepackage{amsmath,amssymb,amsfonts,amsthm,mathtools}
\usepackage{booktabs}
\usepackage{csquotes}
\usepackage[noend]{algorithm, algpseudocode}
\floatname{algorithm}{Algorithm}

\usepackage{graphicx}
\usepackage{textcomp}
\usepackage{flushend}
\usepackage[x11names,svgnames,dvipsnames,table,rgb]{xcolor}
\usepackage[textsize=tiny,loadshadowlibrary,shadow]{todonotes}
\usepackage[colorlinks = true, linkcolor=blue, citecolor=blue, urlcolor  = blue]{hyperref}
\usepackage{braket}
\usepackage{arydshln}
\usepackage{multirow}
\usepackage{float}
\usepackage{color}
\usepackage{soul}
\usepackage{siunitx}
\usepackage{tabularx}
\usepackage{subcaption}
\usepackage{orcidlink}
\usepackage{comment}

\usepackage{tikz}
\usepackage[compat=0.6]{yquant}
\useyquantlanguage{groups}
\usepackage[capitalize]{cleveref}

\usepackage[
	style=ieee,
        citestyle=numeric-comp,
	sortcites=true,
	url=true,
	eprint=true,
	giveninits=false,
	minnames=1,
	maxnames=3
	]{biblatex}
\crefname{equation}{Eq.}{equations}
\crefname{section}{Sec.}{sections}
\crefname{figure}{Fig.}{figures}
\crefname{appendix}{Appendix}{appendices}
\crefname{table}{Table}{tables}

\begin{document}
\title{Quantum Entrepreneurship Lab: \\ Training a Future Workforce \\ for the Quantum Industry}
\author{
\IEEEauthorblockN{%
  Aaron Sander\IEEEauthorrefmark{1}\IEEEauthorrefmark{4}\thanks{Corresponding author: \href{mailto:aaron.sander@tum.de}{aaron.sander@tum.de}}, 
  Rosaria Cercola\IEEEauthorrefmark{2}\IEEEauthorrefmark{5},
  Andrea Capogrosso \IEEEauthorrefmark{3},
  Stefan Filipp\IEEEauthorrefmark{2}\IEEEauthorrefmark{4}\IEEEauthorrefmark{6}, \\[0.5ex]
  Bernhard Jobst\IEEEauthorrefmark{2}\IEEEauthorrefmark{4},
  Christian B. Mendl\IEEEauthorrefmark{1}\IEEEauthorrefmark{4}, 
  Frank Pollmann\IEEEauthorrefmark{2}\IEEEauthorrefmark{4}, 
  Christopher Trummer\IEEEauthorrefmark{5}, \\[0.5ex]
  Isabell Welpe\IEEEauthorrefmark{1},
  Max Werninghaus\IEEEauthorrefmark{2}\IEEEauthorrefmark{4}\IEEEauthorrefmark{6}\IEEEauthorrefmark{7}, 
  Robert Wille\IEEEauthorrefmark{1}\IEEEauthorrefmark{4}\IEEEauthorrefmark{8}\IEEEauthorrefmark{9}
  Christian Wimmer\IEEEauthorrefmark{3}
}
\\
\IEEEauthorblockA{\IEEEauthorrefmark{1}School of Computational, Information, and Technology, Technical University of Munich}
\IEEEauthorblockA{\IEEEauthorrefmark{2}School of Natural Sciences, Technical University of Munich}
\IEEEauthorblockA{\IEEEauthorrefmark{3}School of Management, Technical University of Munich}
\IEEEauthorblockA{\IEEEauthorrefmark{4}Munich Center for Quantum Science and Technology (MCQST)}
\IEEEauthorblockA{\IEEEauthorrefmark{5}TUM Venture Labs Quantum / Semicon}
\IEEEauthorblockA{\IEEEauthorrefmark{6}Walther-Meißner-Institut, Bavarian Academy of Sciences and Humanities}
\IEEEauthorblockA{\IEEEauthorrefmark{7}PeakQuantum GmbH}
\IEEEauthorblockA{\IEEEauthorrefmark{8}Munich Quantum Software Company}
\IEEEauthorblockA{\IEEEauthorrefmark{9}Software Competence Center Hagenberg}
}

\maketitle

\begin{abstract}
The Quantum Entrepreneurship Lab (QEL) is a one-semester, project-based course at the Technical University of Munich (TUM), designed to bridge the gap between academic research and industrial application in the quantum sector.
As part of the Munich Quantum Valley (MQV) ecosystem, the course fosters interdisciplinary collaboration between technical and business students, equipping them with the skills necessary to contribute to or lead in the emerging quantum industry. The QEL curriculum integrates two complementary tracks. First, technical students form teams where they engage in cutting-edge, industry-relevant research topics under academic supervision. Meanwhile business students in a parallel course explore commercialization strategies, risks, and opportunities within the quantum technology landscape. Midway through the semester, a selection of the business students join the technical course to form interdisciplinary teams which assess the feasibility of transforming scientific concepts into viable business solutions. The course culminates in three key deliverables: a publication-style technical report, a white paper analyzing the business potential and financial requirements, and a startup pitch presented to the quantum community at a Demo Day. This work outlines the course structure, objectives, and outcomes, providing a model for other institutions seeking to cultivate a highly skilled, innovation-driven workforce in quantum science and technology.
\end{abstract}
\begin{IEEEkeywords}
	Quantum Technology, Entrepreneurship, Deep Tech, Education
\end{IEEEkeywords}

\section{Introduction}
Quantum technology is rapidly becoming a reality with the 2020s seeing significant investment from both the public and private sectors \cite{jurczak_investing_2023, mckinsey2024quantum, putranto_deep_2024, sotelo_quantum_2021}. From relatively mature technological solutions such as quantum-inspired computational techniques \cite{altmann_quantum-inspired_2018, peddinti_quantum-inspired_2024} to the rapidly evolving quantum computing field \cite{preskill_quantum_2018, deshpande_assessing_2022}, a transformative industry is developing with disruptive innovation across science and engineering. While we can see this on the horizon, the technology is slow to mature and we do not necessarily know what form these industrial solutions will eventually take \cite{martens_acceptance_2025}.

We see this as a motivator to foster an environment which can align talented and ambitious students from across disciplines such that we develop business-minded quantum talent as well as quantum-familiar business talent who can successfully bring academic ideas into something tangible and industry-relevant. More precisely, we see a gap both in the ability for excellent quantum talent to bring their ideas out of academic thinking as well as for business students to broadly apply their knowledge in a deep tech field, where the future is not certain.

To achieve this, we have developed the \emph{Quantum Entrepreneurship Lab (QEL)}, a multi-disciplinary, one semester course at the \emph{Technical University of Munich (TUM)} within the larger \emph{Munich Quantum Valley (MQV)} ecosystem. This is a project-based course split into a technical and a parallel business sub-course. During the course, the technical students are assigned to groups and work on industry-relevant research topics while being coached by academics at TUM, meanwhile the business students learn about the quantum industry including its needs, risks, and possibilities. Halfway through the semester, these courses are combined such that each technical group works with business students who aim to analyze the feasibility of transforming these academic topics into a product. Finally, each team is expected to produce a technical report in the style of a scientific publication, a white paper which gives an honest assessment of the idea's ability to become a product and eventually support a business, as well as a startup pitch-style presentation at a final \emph{Demo Day} in which we invite academics, industry partners, and investors from within our network. Overall, this is intended to mimic a realistic startup journey within deep tech, wherein a business-minded researcher spins out their idea from academia.

We see this as an opportunity to bring students into research and deep tech as early as possible and to develop solid foundations before they begin future jobs, PhDs, or entrepreneurial endeavors. Additionally, for many it becomes one of the first true experiences in a multi-disciplinary team, spanning from business to engineering and physics, a step which we expect many students to encounter in their career, although one which we can prepare them for in advance. This course is also intended to form the basis of a pipeline by which well-rounded talent can enter the quantum industry, whether that be contributing to one of the many quantum companies that exists or developing their own startup. Additionally, it creates a safe environment to try new ideas and evaluate them honestly. Finally, it connects many talented students in the MQV ecosystem with academic groups and industrial partners, stakeholders who can benefit from the research output of the course and the exploratory business models the teams propose. 

This work is intended to describe the environment from which this course was created, its design, its successes, and its challenges. We seek to provide the blueprint such that other institutions can found similar courses, while also inviting feedback from the community on what we can do to improve in our own environment.

\section{What institutions are involved?}
This section provides an overview of the various players in the development of the QEL. Each entity serves as a pillar to create the support needed to mitigate the complexity in creating an impactful course which realizes our learning goals for the students.

\subsection{Munich Quantum Valley}
The Munich Quantum Valley (MQV) is an ecosystem intended to promote quantum technologies, connect researchers with industry, and ultimately, build a full-stack quantum computer in the German state of Bavaria \cite{MQV2024strategy}. This provides both a funding structure and cultivates a cooperative environment to connect quantum-related groups within Munich. The MQV is key for providing the network and connections between Munich's universities, established industry with interest in quantum, as well as new quantum companies. The MQV is a collaboration between several institutions, in particular, the Technical University of Munich,  Ludwig-Maximilians Universität München (University of Munich, LMU), Friedrich-Alexander University (FAU), Max Planck Institute of Quantum Optics (MPQ), Fraunhofer Institutes, and the German Aerospace Center (DLR) such that it provides access to a full quantum network. The QEL actively uses this collaborative network for funding, project ideas, as well as a channel for both incoming and outgoing quantum talent.

\subsection{Technical University of Munich}
The Technical University of Munich (TUM) is one of Germany's (and Europe's) elite universities which provides the talent needed for the QEL. Under the motto "The Entrepreneurial University", the QEL pulls heavily from its Master's in Quantum Science and Technology (in combination with LMU), Management and Technology, as well as related studies for its participants. Additionally, the course organizers and team coaches come from the academic groups within its various faculties, ranging from the School of Management to the School of Natural Sciences and the School of Computation, Information, and Technology. The QEL relies on TUM for providing students, industry-relevant project ideas, team coaches, and course organization.

\subsection{TUM Venture Lab Quantum / Semicon}
TUM Venture Lab Quantum / Semicon (VLQS), is a start-up incubator with a focus on quantum and semiconductor technologies, part of the TUM Venture Labs, an initiative by TUM and UnternehmerTUM. VLQS provides support for researchers, including the network and infrastructure needed to turn academic research into startups. Within this hub are several Munich-based quantum startups, particularly in the space of quantum hardware. The QEL relies on VLQS for both the course organization, lecture rooms, and follow-up support for the teams after conclusion of the course. This includes additional programs, ongoing advisory support, office space, and connections with venture capital that the students may use for furthering their ideas. 

\subsection{PushQuantum e.V.}
PushQuantum e.V. (PQ) is a student initiative from which the original idea for the QEL was created and executed. This club serves as an environment for educating students across fields on quantum technology as well as providing networking opportunities to support their future careers. Initially based around the QEL, PQ is now a much larger organization which covers industry networking, workshops, and outreach events for students interested in the quantum field. PQ provides the QEL with student volunteers who assist with the organization, marketing, and IT support necessary for running the course. Its leadership ensures the sustainability and continuity of the course between semesters.

\begin{figure}[h!]
    \centering
    \includegraphics[width=\linewidth]{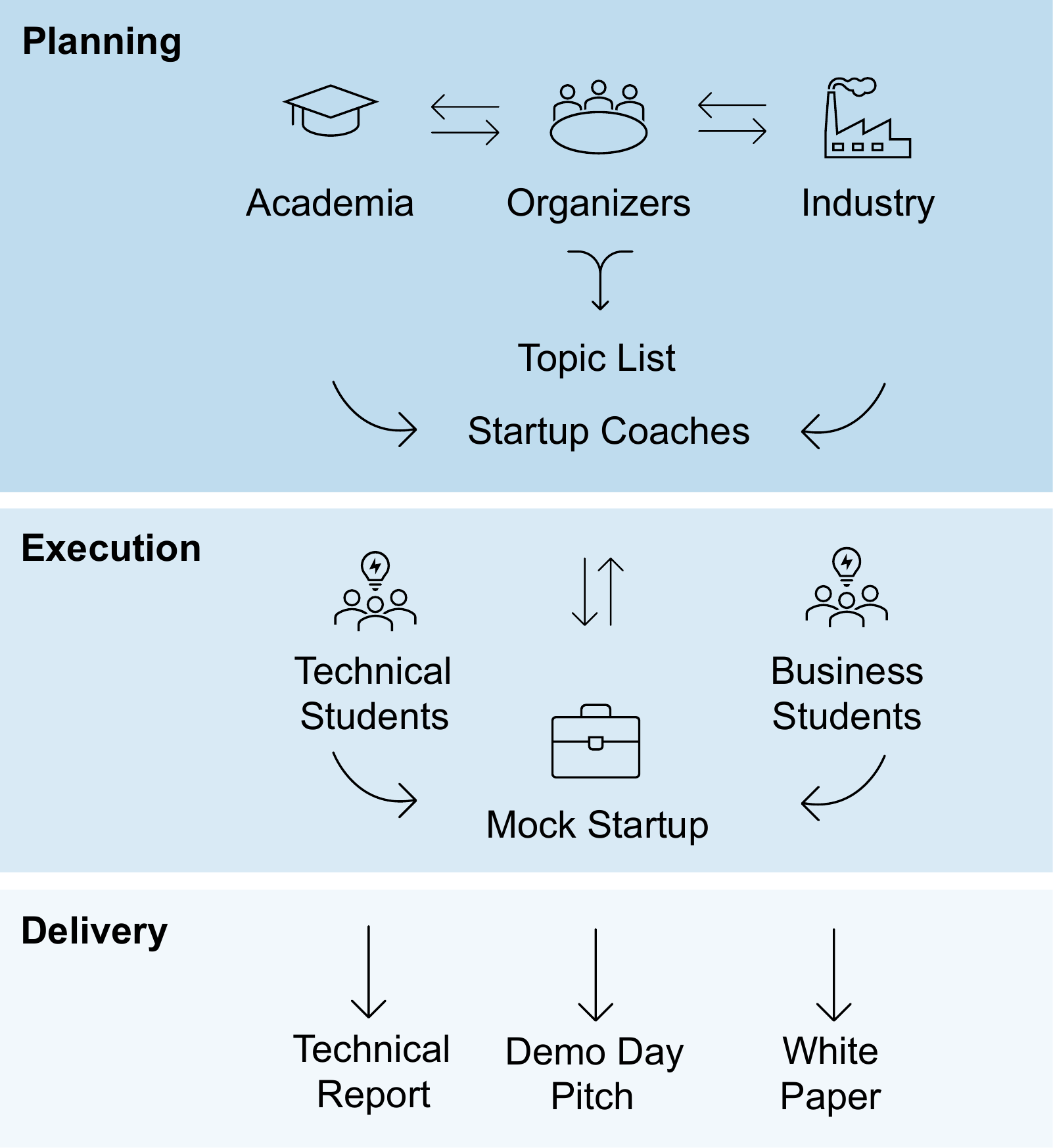}
    \caption{Key components and organizational structure of the Quantum Entrepreneurship Lab}
    \label{fig:Organization}
\end{figure}

\section{Course Design}
This section outlines the overall course structure and the primary elements that contribute to its teaching goals. The course is delivered over one academic semester at the Technical University of Munich--approximately six months of time--with four months of active lectures and a subsequent two-month period dedicated to exams or project deliverables. The design discussed in this section is visualized in \cref{fig:Organization}.

The format described here reflects the QEL design implemented from April to October 2024, acknowledging that it is neither the original structure nor the final iteration. In particular, it consists mostly of a technical course with business roles joining at the midway point. The upcoming iteration will include a fully-integrated sister course from which students will be selected to merge into the course structure described here. We elucidate these changes in \cref{sec:Improvements}.

\subsection{Objectives}
The QEL is driven by several key objectives that benefit students, academic groups involved in the course, Munich’s quantum industry, and the quantum sector at large. Addressing the diverse needs of these stakeholders requires meticulous planning, ongoing supervision, and targeted guidance throughout every phase of the course. In essence, the QEL aims to achieve the following:

\subsubsection{Introducing students to research early}
Early engagement in research is expected to produce more impactful outcomes in students’ future careers, whether they pursue academic research or industry roles \cite{rodenbusch2016, bowman_getting_2018}. The QEL offers an opportunity for students to acquire and apply skills beyond traditional classroom settings by engaging with genuine research topics, thereby allowing them to explore their interests in a structured yet innovative environment.

\subsubsection{Introducing students to deep tech}
Deep tech \cite{mit2023deeptech} encompasses cutting-edge fields such as quantum computing, artificial intelligence, space technology, and biotechnology. These industries face significant challenges and offer transformative potential; however, the nascent state of these technologies often complicates the development of concrete ideas and business models. The QEL serves as a direct entry point into the deep tech landscape, acting as an essential stepping stone for translating innovative concepts into practical solutions.

\subsubsection{Producing research output for academic groups}
The course utilizes legitimate research topics sourced from the academic groups associated with the program. Whether the projects are active initiatives or exploratory ideas, the QEL provides a collaborative platform to advance these topics. In optimal cases, student work may result in publishable outcomes that benefit both the researchers and the students, while also fostering connections that may lead to future employment and collaborative opportunities.

\subsubsection{Interdisciplinary Collaborations}
The QEL fosters a dynamic, interdisciplinary learning environment by bringing together students from diverse backgrounds, including physics, computer science, engineering, and management. This cross-disciplinary approach enables participants to tackle complex quantum-related challenges. Such collaboration not only enriches the learning process but also mirrors the structure of real-world quantum technology ventures, where success depends on a seamless blend of scientific innovation and strategic implementation.

\subsubsection{Industry Involvement and Mentorship}
A key pillar of the QEL is its strong connection to industry experts and professionals in the quantum technology sector in the Munich Quantum Valley. Throughout the course, students benefit from guest lectures, hands-on workshops, and direct mentorship from experienced researchers, entrepreneurs, and industry leaders. These interactions provide relevant insights into the latest advancements, market trends, and challenges in the field. Furthermore, industry engagement facilitates networking opportunities that can lead to internships, job placements, or even investment opportunities for student-led ventures. This collaboration ensures that students not only develop technical and research skills but also gain an understanding of the business and entrepreneursial  landscape and commercial viability of their ideas.

\subsubsection{Practical Prototyping and Proof of Concepts}
Beyond theoretical learning, the QEL emphasizes hands-on experimentation of quantum technologies. Students are encouraged to translate research into tangible prototypes or proof-of-concept demonstrations. Whether through quantum algorithms, hardware development, or quantum-inspired solutions, participants work on projects that push the boundaries of feasibility and innovation in a practical sense. Access to lab resources, computational tools, and expert guidance ensures that students can experiment with their concepts, refine their methodologies, and ultimately create industry-relevant solutions that have the potential to transition into further research, patents, or startups.

\subsubsection{Creating a quantum workforce}
The QEL is designed to cultivate the practical application of the academic knowledge that both technical and business students acquire during their studies. By promoting interdisciplinary teamwork, the course ensures that students develop the ability to communicate effectively across diverse fields such as business, engineering, and physics. These collaborative skills are crucial for addressing the complex challenges of the quantum industry and ultimately contribute to building a robust, competent quantum workforce.

\subsection{Selecting industry-relevant topics}
A core principle of the QEL's design is that students work on quantum topics with direct industry relevance. In our context, "industry-relevant" refers to near-term applications—projects that address immediate market needs or fill existing gaps rather than long-term speculative innovations. For this iteration of the course, we have identified two overlapping categories:

\subsubsection{Quantum-inspired techniques}
This category focuses on advanced computational methods, such as tensor networks, machine learning algorithms, and optimization techniques, that originated from quantum science, often developed in parallel with the advent of quantum computers. Traditionally applied in physics, these techniques are now being explored in machine learning \cite{huynh_quantum-inspired_2023}, fluid dynamics \cite{peddinti_quantum-inspired_2024}, generative AI \cite{moussa_application_2023}, and data science \cite{cichocki_tensor_2014} demonstrating their potential to drive innovation across multiple fields.

\subsubsection{Solutions for the quantum industry}
Rather than viewing the quantum industry solely as a collection of competing entities, we see it as a collaborative ecosystem working toward common objectives, such as achieving a stable quantum computer. Amid significant investment, there exists a substantial opportunity to develop products and services for quantum companies, for example, tailor-made simulators \cite{xu_herculean_2023}, compilers \cite{maronese_quantum_2021}, or consulting on error mitigation and correction strategies \cite{filippov_scalability_2024}.

By focusing on these two categories, the QEL ensures that student projects not only tackle real-world challenges but also contribute to the evolving landscape of the quantum industry. However, we expect that the topics in this course will fluctuate with the changing needs of the quantum landscape, in particular, including more hardware-focused topics, hybrid classical-quantum computing \cite{yang_semiconductor_2022, ovalle_magallanes_hybrid_2022}, and applications of quantum computing \cite{bova_commercial_2021}.
In the end, the following topics were provided:
\begin{enumerate}
    \item Quantum Compiler Development
    \item Resource Estimation of Quantum Algorithms
    \item Quantum-inspired Generative AI
    \item Data Generation for Quantum Machine Learning
    \item Machine Learning for Quantum Chemistry
    \item HPC Integration of Quantum Computers
    \item Development Tools for Quantum Computing
    \item Quantum-inspired Data Science
    \item Quantum Hardware/Error Correction Codesign
\end{enumerate}

\subsection{Selecting participants}
The course targets Master's-level students who are typically in their 2nd semester (out of a typical 4 semesters in the German university system). For many studies at TUM, this directly precedes a year-long Master's thesis such that the QEL can serve as a preparation for more successful and impactful work. For the QEL to deliver on its objectives, we most importantly need motivated students who can successfully make progress on the topics. For this reason, we limit the participation to 5 teams of 5, i.e., 15 technical and 10 business students, who are selected through an application process. Each student sends a CV as well as basic information about themselves with the option to send a short clarification about their CV if needed. In the selection process, we prioritize several factors and de-prioritize others by assigning ratings to several attributes of the participants. This is intended to maximize motivation, reliability in participation, and technical skill. The selection process, as outlined below, does not prevent students with a technical background from assuming a business role and (in exceptional cases) vice versa. Therefore we will refer to them as technical and business roles.

\subsubsection{Technical roles}
First, we rate the focus on quantum science in their academic major. Students who have an entirely quantum-focused major receive the most points. In practice, this prioritizes students in Quantum Science and Technology, Physics or Math with a concentration in quantum science, and Computer Science with experience in quantum computing. Next, we rank the time remaining in their degree such that students with more experience as well as those who are repeat applicants are prioritized. Finally, we rate their industry and entrepreneurship experience where we prioritize any industry experience, but especially experience in the quantum industry or startups in general. These factors help guarantee that students are both experienced and motivated. Notably, we purposely do not look at grades as these are not useful for forecasting entrepreneurial success. Furthermore, students who may have worked during their studies can very likely have worse grades, yet invaluable experience and business acumen.

\subsubsection{Business roles}
Here, we prioritize the fit of extracurricular activities. As with the technical roles, those with previous work experience in the quantum industry are prioritized strongest, followed by general startup experience. Students who had entrepreneurial experience, either through clubs or even those who have been involved with founding their own startups previously are also heavily prioritized. Generally, we see that most students fulfilling these criteria come from Management and Technology, Management, and Consumer Science, although we did not explicitly prioritize any specific degree.

\subsubsection{Diversity and Inclusion}
Recognizing the importance of diverse perspectives in quantum entrepreneurship, the QEL actively encourages participation from students with varied backgrounds. The program seeks to create a balanced cohort by ensuring representation from both technical and business disciplines, fostering an environment where interdisciplinary collaboration thrives. Gender diversity is also a key consideration, as the quantum industry and startup ecosystem have historically been male-dominated. Efforts are made to recruit and support underrepresented groups in STEM and business fields, helping to build a more inclusive quantum ecosystem. Additionally, international students and those from non-traditional academic backgrounds—such as law, ethics, or policy in quantum technology—are considered, particularly if their experience aligns with the program’s objectives of bridging deep tech and entrepreneurship.

\begin{figure*}
    \centering
    \includegraphics[width=0.7\linewidth]{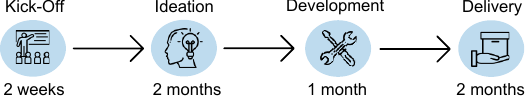}
    \caption{Overall course timeline described in \cref{sec:Structure}}
    \label{fig:Timeline}
\end{figure*}

\subsection{Structure} \label{sec:Structure}
The course follows a structured, multi-phase approach designed to simulate the startup journey. The structure described below reflects the previous iteration of the course and is visualized in \cref{fig:Timeline}. Changes for the upcoming version, particularly regarding business components, will be detailed in \cref{sec:Improvements}.

\subsubsection{Kick-off phase (2 weeks)}
The course begins with a \textbf{Welcome Day} for the technical roles, outlining expectations and introducing potential topics. This session provides an informal setting for students to network and discuss their interests. Over the next few days, students rank their topic preferences and indicate preferred teammates to foster team cohesion. Topics are then assigned accordingly.

Each team then meets with the academic group responsible for their topic and is introduced to a \textbf{team coach}, ideally a PhD student working in the field or an employee of a quantum company, who provides research guidance. Coaches meet with teams \textbf{at least biweekly} (every two weeks) to monitor progress, introduce research methodologies, and ensure steady advancement toward realistic goals.

\subsubsection{Ideation phase (2 months)}
After topic selection, teams begin \textbf{planning} and \textbf{exploring} their research problem. This phase is designed to help students familiarize themselves with each other’s working styles and establish regular meeting schedules. To broaden their understanding, students attend lectures covering all course topics, ensuring exposure beyond their own specialization.

Critical thinking is emphasized, particularly in the context of the quantum field, which is vast and often subject to hype. Students are encouraged to question assumptions, recognizing that:
\begin{itemize}
    \item Provided topics are not inherently perfect.
    \item Uncovered topics may still be viable.
    \item Success in research is never guaranteed, and open discussions with collaborators (i.e., peers, coaches, and organizers) are key to shaping the project’s direction.
\end{itemize}
To guide their analysis, teams consider questions such as:
\begin{itemize}
    \item \textit{Is the idea too simple or overly complex?}
    \item \textit{Is the research field unexplored or oversaturated?}
    \item \textit{Is the idea outdated or too long-term to be practical?}
\end{itemize}

During this phase, each team delivers \textbf{two progress report presentations}. The first is an informal update, covering their current understanding, research steps, and key discussions. The second, which coincides with the \textbf{Welcome Day for business roles} requires technical roles to distill their topic for a non-technical audience. At this point, business roles join the course and engage with these presentations, ask questions, and subsequently rank the teams they would like to join.

\subsubsection{Development phase (1 month)}
At this stage, combined technical-business teams work toward a final technical result and explore commercialization pathways. Business roles lead efforts in business model development, roadmaps, and branding. This phase is supported by lectures from industry professionals who have successfully transitioned research into commercial products as well as on startup ecosystems and venture capital strategies.

The phase concludes with a \textbf{final progress report}, where teams provide a critical assessment of the technical viability of their research, the potential business case, and realistic requirements for transitioning the idea into an industrial solution. Rather than pitching their project, students evaluate what is needed to launch a startup, including resource estimates, funding requirements, and a roadmap to achieving financial sustainability.

\subsubsection{Delivery phase (2 months)}
In this final phase, students consolidate their work into three key deliverables that represent the course's primary outcomes. 

Early in the phase, students participate in a \textbf{Demo Day}, delivering an \textbf{8--10 minute startup pitch} to an audience of academics, quantum industry professionals, and members of our local network. The primary objective is to advocate for funding, seek collaborators, and engage with potential industry partners. This experience not only introduces students to the art of startup pitching but also enhances their ability to communicate complex ideas in an accessible manner. Additionally, the Demo Day serves as a valuable networking event, helping students connect with industry experts, explore internship opportunities, and establish relationships that could support their careers in the quantum sector.

After Demo Day, the lecture period at the Technical University of Munich concludes, and students are given a flexible timeframe to submit their final reports until the end of the semester. The \textbf{technical report} is expected to follow a publishable format, typically resembling APS or IEEE styles, though students may select an alternative format if more suitable for their research goals. 

In parallel, business roles produce a \textbf{white paper}. The document first translates the technical topic into a form accessible to a non-specialist audience and then critically assesses its commercial viability. This includes an analysis of risks and rewards associated with further investment, an exploration of potential business models, and an evaluation of branding strategies. Additionally, the report examines market competition, identifies target niches, and provides a realistic projection of the financial resources required to scale the idea into a sustainable business. 

Together, these deliverables ensure that students not only develop their technical and research skills but also gain exposure to the practical challenges of transforming an academic idea into an industry-ready innovation.

\begin{figure*}
    \centering
    \includegraphics[width=\linewidth]{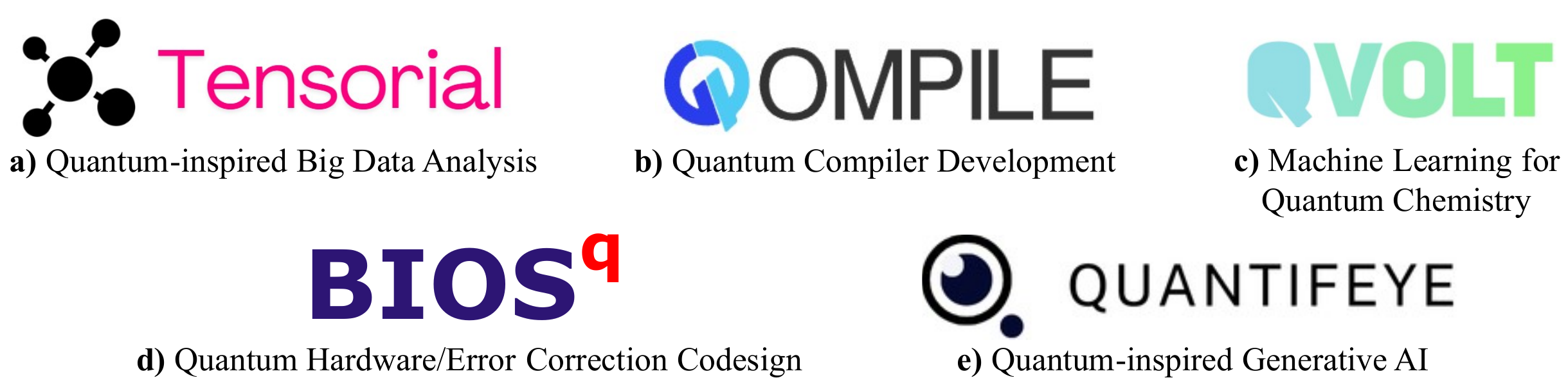}
    \caption{Mock startups created from the selected topics during this iteration of the QEL}
    \label{fig:Projects}
\end{figure*}

\section{Outcomes}
The QEL provides a unique setting for students to explore the intersection of quantum technology and business innovation. While its impact on students and their community can be difficult to quantify, this section highlights key takeaways, particularly student achievements and notable success stories following the course. Each subsection presents one of the five topics selected by students, followed by a summary of post-course impact. The mock startups and their associated topics are shown in \cref{fig:Projects}.

\subsection{Quantum-inspired data science}
This project explored the use of tensor networks for data compression, addressing the increasing computational and energy demands of artificial intelligence. The team developed a quantum-inspired approach using matrix product states (MPS) \cite{orus_practical_2014} to compress the required storage space of data sets, with potential applications in optimizing machine learning models, particularly large-scale neural networks.

On the business side, the team examined the potential for offering software solutions focused on AI model compression. They identified key industry challenges, such as scalability and market adoption, and sought validation for their approach through partnerships and industry engagement.

\subsection{Quantum compiler development}
Students in this project focused on optimizing quantum compilation for superconducting qubits \cite{arute_quantum_2019}, with the goal of improving algorithm performance on noisy intermediate-scale quantum (NISQ) devices \cite{preskill_quantum_2018}. Their approach aimed to reduce computational overhead and enhance resilience against noise through specialized compiler techniques.

The team explored the commercial potential of customized compilation solutions for quantum hardware providers. While recognizing challenges such as pricing and competition from free alternatives, they saw value in tailored optimizations that could accelerate research workflows.

\subsection{Machine learning for quantum chemistry}
This project applied machine learning to accelerate quantum chemistry simulations, particularly in the context of materials discovery. The students developed models to streamline key computational processes, such as band structure calculations and Green’s function predictions, which are crucial in material science research \cite{kuramoto_quantum_2020, sheridan_data-driven_2021}.

From a business perspective, the team considered applications in the development of alternative materials for industrial use. They discussed the potential for collaborations with established companies in the field, but also acknowledged the high investment and long-term validation required for adoption.

\subsection{Quantum-inspired generative AI}
Leveraging techniques from quantum many-body physics, this project explored the application of matrix product states (MPS) \cite{orus_practical_2014} to generative AI models, with a focus on financial applications \cite{mugel_use_2020}. The students investigated how these models could improve predictive capabilities in areas such as stock and option pricing.

The team evaluated the feasibility of commercializing their approach, identifying opportunities in financial modeling and risk assessment. They noted barriers related to technical complexity and industry adoption but saw potential for quantum-inspired AI in specialized financial markets.

\subsection{Quantum hardware/error correction codesign}
This project examined quantum error correction (QEC) strategies tailored to different hardware architectures \cite{roffe_quantum_2019}. The students developed methods to adapt QEC techniques based on the specific characteristics of various quantum platforms, such as superconducting qubits \cite{arute_quantum_2019}, trapped ions \cite{cirac_quantum_1995}, and neutral atoms \cite{wintersperger_neutral_2023}.

The team explored potential industry applications, particularly in partnerships with quantum hardware developers. They acknowledged the long timeline for quantum computing commercialization and the engineering challenges associated with implementing scalable QEC solutions.

\subsection{Student achievements and impact}
The QEL has led to several notable outcomes for participating students. Many teams continued developing their projects beyond the course, with multiple technical contributions resulting in research publications. Some students secured internships or industry collaborations based on connections made through the program, particularly during the Demo Day. Finally, nearly half of the participants continued working within quantum entrepreneurship through follow-up fellowship programs which facilitate project work with various industry partners in the MQV environment.

For teams considering commercialization, the course provided a foundation for exploring market opportunities and challenges. Several groups engaged with industry professionals to refine their business strategies, and one team is actively pursuing a startup. The program also fostered interdisciplinary collaboration, equipping students with both technical and entrepreneurial skills applicable to future careers in quantum technology and beyond.

\subsection{Collaboration with Industry and Research Institutions}
A significant outcome of the QEL is the establishment of collaborations between student teams and industry partners, research institutions, and investors. Some mock startups originating from the program will (hopefully) continue as early-stage ventures, exploring seed funding and incubator support. Additionally, partnerships initiated during the QEL should lead to internships, research collaborations, and mentorship opportunities, bridging the gap between academia and industry. The iterative nature of the course ensures that each cohort refines and expands upon previous innovations, strengthening the long-term impact of the program on the broader quantum ecosystem.

\section{Challenges} \label{sec:Challenges}
Overall, we see the course as a success due to its significant hit rate for several desired outcomes, between academic projects, internships, and general knowledge gained. Despite this, the course design leads to several challenges which we elucidate in this section. This serves to pass on the lessons that we learned, as well as to justify future improvements we plan to make. Generally, our goal is to make this course sustainable such that it does not need to be re-designed or re-organized from the beginning for each iteration.

\subsection{Finding topics}
Selecting suitable topics for student projects requires balancing several key factors: they should be feasible for students to complete within the course timeline, well-suited for collaborative work, sufficiently research-oriented, and open-ended enough to encourage exploration, yet structured enough to provide clear direction. In the quantum field, this balance is particularly challenging due to the rapidly evolving nature of the industry.

While some topics may seem like obvious choices, there is a risk that widely recognized areas have already been extensively explored, diminishing their novelty and impact. This follows a pattern similar to investment trends, i.e., once a technology becomes mainstream knowledge, the opportunity for groundbreaking contributions may have already passed. Careful topic selection, therefore, requires foresight, industry awareness, and an understanding of emerging research directions.

\subsection{Weak links}
One of the core problems with group projects is of course, weak links in the team structure. In this case, we mean not only people who do not pull their weight, but also students who fully drop out of the course. Due to the application process, typically dropping out is much more common than a lack of effort, yet also more problematic. Generally because of the nature of our projects, there is enough redundancy in the number of students that this can be mitigated if a single student drops out of the team. However, especially in previous iterations, we have had multiple students leave such that teams had to be redistributed.

In order to mitigate this, we plan to implement a waiting list such that students can continue to apply through the ideation phase. While this is a simple solution, we expect this to be the strongest support, especially if the number of applicants can be kept sufficiently high.

\subsection{Coaches}
Ideally, each topic is guided by a dedicated coach whose expertise aligns closely with the teams’ needs. In this setup, each coach works with one team, meeting biweekly in a way that balances well with their primary responsibilities. While this structure works well in principle, securing a sufficient number of coaches with the right expertise can be a challenge.

The main difficulty arises from two factors, one logistical and one structural. First, many qualified coaches have demanding schedules, making it difficult for them to commit, particularly for a course that relies on voluntary participation. Second, the pool of academics who are both highly specialized and industry-minded is naturally limited. However, creative solutions such as engaging postdocs, industry professionals, or alumni of the course can help bridge this gap.

\section{Future Improvements} \label{sec:Improvements}
\subsection{Re-integrating industry partners}
In the initial design of the QEL (2020–2023), industry partners directly provided project topics rather than academic groups. This approach was intended to strengthen connections between students and potential collaborators or employers. However, it required coordinating multiple external stakeholders each semester, making it difficult to ensure consistent participation. In some cases, companies were unable to support the course due to time constraints or internal policies leading to a shortage of topics and, in extreme cases, course cancellations.

In the upcoming iteration, we plan to reintroduce industry engagement by allowing companies to submit both project topics and coaches. Unlike previous editions, the course will now have a stable foundation of academic topics, ensuring continuity while maintaining opportunities for industry collaboration. This hybrid approach provides a more sustainable model, balancing academic rigor with real-world relevance.

\subsection{Improving business education}
In the previous format, business roles joined the course at the midterm point of the semester. This structure was chosen because the initial phase focuses on technical roles developing a deep understanding of their topics, leaving little direct work for business roles early on. While this approach is preferable to having them join at the beginning without meaningful engagement, it presents certain challenges.

First, mid-semester enrollment is not particularly attractive to many students, as their schedules are typically set by that point, making it difficult to commit to an additional project-based course. Second, this format limits our ability to provide business students with a strong foundation in quantum computing before they integrate with the teams.

To address these issues, future iterations of the course will include a parallel business-focused course on the quantum industry. This course will run for the full semester, allowing students to gain relevant knowledge and engage in structured learning before merging with the technical teams through a selective application process. This revised structure ensures that business students receive adequate preparation, makes the course more appealing for credit and scheduling purposes, and fosters a more seamless integration with the Quantum Entrepreneurship Lab.

\subsection{Expanding Networking and Mentorships }
One of the key advantages of the QEL is its potential to connect students with industry leaders, researchers, and entrepreneurs in the quantum technology sector. To further enhance these opportunities future iterations of the course will introduce even more networking events, panel  discussions, and exchange programs. Additionally, a structured and innovative hybrid mentorship framework will be established from industry experts and past QEL alumni.

\subsection{Entrepreneurs of Tomorrow}
While the QEL provides students with exposure to both quantum technology and business innovation, there is an opportunity to further develop their entrepreneurial mindset and skills. Beyond traditional learning, students will immerse themselves in quantum startup simulation labs, where they will prototype next-gen quantum applications, stress-test business models against future market scenarios, and engage in AI-assisted venture forecasting. Real-world case studies will be transformed into interactive, challenge-based sprints, allowing students to tackle the same hurdles faced by today’s quantum pioneers. 

\section*{Acknowledgments}
We would like to thank Raúl Berganza Gómez and Tom Hubregtsen for laying the foundation of the original Quantum Entrepreneurship Lab, as well as Benjamin Schiffer for his insights.

We would also like to acknowledge the participants of the course including 
Egor Agapov, Oriol Bertomeu, Andrés Carballo, Rhea Daga, Raunaq Jain, Danylo Kolesnyk, Jessica Link, Alexander Orlov, Daniel Ortmann, Mert Özel, Alan Piovesana, Max Gert Richter, Leon Seidelmeier, Agata Skoczylas, Laurin Steiner, Andreas Strobl, Myron Theocharakis, Dimitrios Vasileiadis, and Yelyzaveta Vodovozova.

This work has been supported by the Munich Quantum Valley, which is supported by the Bavarian state government with funds from the Hightech Agenda Bayern Plus. Additionally, the authors have received funding from the European Research Council (ERC) under the European Union’s Horizon 2020 research and innovation program (grant agreement No.\ 101001318) and the Deutsche Forschungsgemeinschaft (DFG, German Research Foundation) under Germany’s Excellence Strategy–EXC–2111–390814868.


\printbibliography

@techreport{mckinsey2024quantum,
  author    = {M. Bogobowicz and K. Dutta and M. Gschwendtner and A. Heid and M. Issler and N. Mohr and H. Soller and R. Zemmel and A. Zhang},
  title     = {Steady Progress in Approaching the Quantum Advantage},
  institution = {McKinsey \& Company},
  year      = {2024},
  month     = {April},
  note      = {Accessed: 2025-02-21, \url{https://www.mckinsey.com/capabilities/mckinsey-digital/our-insights/steady-progress-in-approaching-the-quantum-advantage}}
}

@article{preskill_quantum_2018,
  author = {J. Preskill},
  title = {Quantum Computing in the NISQ era and beyond},
  journal = {Quantum},
  volume = {2},
  pages = {79},
  year = {2018},
  month = {August},
  doi = {10.22331/q-2018-08-06-79},
  url = {http://arxiv.org/abs/1801.00862}
}

@article{sotelo_quantum_2021,
  author = {R. Sotelo},
  title = {Quantum Computing Entrepreneurship and IEEE TEMS},
  journal = {IEEE Eng. Manag. Rev.},
  volume = {49},
  number = {3},
  pages = {26--29},
  year = {2021},
  doi = {10.1109/EMR.2021.3098260},
  url = {https://ieeexplore.ieee.org/document/9492020}
}

@article{deshpande_assessing_2022,
  author = {A. Deshpande},
  title = {Assessing the quantum-computing landscape},
  journal = {Commun. ACM},
  volume = {65},
  number = {10},
  pages = {57--65},
  year = {2022},
  month = {October},
  doi = {10.1145/3524109},
  url = {https://dl.acm.org/doi/10.1145/3524109}
}

@online{jurczak_investing_2023,
  author = {C. Jurczak},
  title = {Investing in the Quantum Future: State of Play and Way Forward for Quantum Venture Capital},
  publisher = {arXiv},
  year = {2023},
  month = {December},
  note = {arXiv:2311.17187 [physics]},
  url = {http://arxiv.org/abs/2311.17187}
}

@article{putranto_deep_2024,
  author = {D. S. C. Putranto and R. W. Wardhani and J. Ji and H. Kim},
  title = {A Deep Inside Quantum Technology Industry Trends and Future Implications},
  journal = {IEEE Access},
  volume = {12},
  pages = {115776--115801},
  year = {2024},
  doi = {10.1109/ACCESS.2024.3444779},
  url = {https://ieeexplore.ieee.org/document/10638045}
}

@techreport{MQV2024strategy,
  author    = {{Munich Quantum Valley Initiative}},
  title     = {Strategiepapier: Munich Quantum Valley Initiative},
  institution = {Munich Quantum Valley},
  year      = {2024},
  note      = {Accessed: 2025-02-21, \url{https://www.munich-quantum-valley.de/}}
}

@article{cirac_quantum_1995,
  author = {J. I. Cirac and P. Zoller},
  title = {Quantum Computations with Cold Trapped Ions},
  journal = {Phys. Rev. Lett.},
  volume = {74},
  number = {20},
  pages = {4091--4094},
  year = {1995},
  month = {May},
  doi = {10.1103/PhysRevLett.74.4091},
  url = {https://link.aps.org/doi/10.1103/PhysRevLett.74.4091}
}

@article{arute_quantum_2019,
  author = {F. Arute et al.},
  title = {Quantum supremacy using a programmable superconducting processor},
  journal = {Nature},
  volume = {574},
  number = {7779},
  pages = {505--510},
  year = {2019},
  month = {October},
  doi = {10.1038/s41586-019-1666-5},
  url = {https://www.nature.com/articles/s41586-019-1666-5}
}

@article{peddinti_quantum-inspired_2024,
  author = {R. D. Peddinti et al.},
  title = {Quantum-inspired framework for computational fluid dynamics},
  journal = {Commun. Phys.},
  volume = {7},
  number = {1},
  pages = {1--7},
  year = {2024},
  month = {April},
  doi = {10.1038/s42005-024-01623-8},
  url = {https://www.nature.com/articles/s42005-024-01623-8}
}

@article{altmann_quantum-inspired_2018,
  author = {Y. Altmann et al.},
  title = {Quantum-inspired computational imaging},
  journal = {Science},
  volume = {361},
  number = {6403},
  pages = {eaat2298},
  year = {2018},
  month = {August},
  doi = {10.1126/science.aat2298},
  url = {https://www.science.org/doi/10.1126/science.aat2298}
}

@article{rodenbusch2016,
  author = {S. E. Rodenbusch and P. R. Hernandez and S. L. Simmons and E. L. Dolan},
  title = {Early Engagement in Course-Based Research Increases Graduation Rates and Completion of Science, Engineering, and Mathematics Degrees},
  journal = {CBE Life Sci. Educ.},
  volume = {15},
  number = {2},
  pages = {ar20},
  year = {2016},
  doi = {10.1187/cbe.16-03-0117},
  url = {https://www.ncbi.nlm.nih.gov/pmc/articles/PMC4909342/}
}

@article{bowman_getting_2018,
  author = {N. A. Bowman and J. M. Holmes},
  title = {Getting off to a good start? First-year undergraduate research experiences and student outcomes},
  journal = {High. Educ.},
  volume = {76},
  number = {1},
  pages = {17--33},
  year = {2018},
  month = {July},
  doi = {10.1007/s10734-017-0191-4},
  url = {https://doi.org/10.1007/s10734-017-0191-4}
}

@techreport{mit2023deeptech,
  author = {F. Murray and P. Budden and J. Guimón},
  title = {What is "Deep Tech" and what are Deep Tech Ventures?},
  institution = {MIT Regional Entrepreneurship Acceleration Program},
  year = {2023},
  note = {Accessed: 2025-02-21, \url{https://reap.mit.edu/assets/What_is_Deep_Tech_MIT_2023.pdf}}
}

@online{huynh_quantum-inspired_2023,
  author = {L. Huynh and J. Hong and A. Mian and H. Suzuki and Y. Wu and S. Camtepe},
  title = {Quantum-Inspired Machine Learning: A Survey},
  publisher = {arXiv},
  year = {2023},
  month = {September},
  note = {arXiv:2308.11269 [cs]},
  doi = {10.48550/arXiv.2308.11269},
  url = {http://arxiv.org/abs/2308.11269}
}

@inproceedings{moussa_application_2023,
  author = {C. Moussa and H. Wang and M. Araya-Polo and T. Bäck and V. Dunjko},
  title = {Application of quantum-inspired generative models to small molecular datasets},
  booktitle = {Proc. IEEE Int. Conf. Quantum Comput. Eng. (QCE)},
  pages = {342--348},
  year = {2023},
  month = {September},
  doi = {10.1109/QCE57702.2023.00046},
  url = {http://arxiv.org/abs/2304.10867}
}

@online{cichocki_tensor_2014,
  author = {A. Cichocki},
  title = {Tensor Networks for Big Data Analytics and Large-Scale Optimization Problems},
  publisher = {arXiv},
  year = {2014},
  month = {August},
  note = {arXiv:1407.3124 [cs, math]},
  doi = {10.48550/arXiv.1407.3124},
  url = {http://arxiv.org/abs/1407.3124}
}

@inproceedings{yang_semiconductor_2022,
  author = {Y.-F. Yang and M. Sun},
  title = {Semiconductor Defect Detection by Hybrid Classical-Quantum Deep Learning},
  booktitle = {Proc. IEEE/CVF Conf. Comput. Vis. Pattern Recognit. (CVPR)},
  address = {New Orleans, LA, USA},
  pages = {2313--2322},
  year = {2022},
  month = {June},
  doi = {10.1109/CVPR52688.2022.00236},
  url = {https://ieeexplore.ieee.org/document/9879978/}
}

@article{ovalle_magallanes_hybrid_2022,
  author = {E. Ovalle-Magallanes and J. G. Avina-Cervantes and I. Cruz-Aceves and J. Ruiz-Pinales},
  title = {Hybrid classical-quantum Convolutional Neural Network for stenosis detection in X-ray coronary angiography},
  journal = {Expert Syst. Appl.},
  volume = {189},
  pages = {116112},
  year = {2022},
  month = {March},
  doi = {10.1016/j.eswa.2021.116112},
  url = {https://www.sciencedirect.com/science/article/pii/S0957417421014421}
}

@article{bova_commercial_2021,
  author = {F. Bova and A. Goldfarb and R. G. Melko},
  title = {Commercial applications of quantum computing},
  journal = {EPJ Quantum Technol.},
  volume = {8},
  number = {1},
  pages = {1--13},
  year = {2021},
  month = {December},
  doi = {10.1140/epjqt/s40507-021-00091-1},
  url = {https://epjquantumtechnology.springeropen.com/articles/10.1140/epjqt/s40507-021-00091-1}
}

@online{xu_herculean_2023,
  author = {X. Xu and S. Benjamin and J. Sun and X. Yuan and P. Zhang},
  title = {A Herculean task: Classical simulation of quantum computers},
  publisher = {arXiv},
  year = {2023},
  month = {February},
  note = {arXiv:2302.08880 [quant-ph]},
  doi = {10.48550/arXiv.2302.08880},
  url = {http://arxiv.org/abs/2302.08880}
}

@online{filippov_scalability_2024,
  author = {S. N. Filippov and S. Maniscalco and G. García-Pérez},
  title = {Scalability of quantum error mitigation techniques: from utility to advantage},
  publisher = {arXiv},
  year = {2024},
  month = {March},
  note = {arXiv:2403.13542 [quant-ph]},
  doi = {10.48550/arXiv.2403.13542},
  url = {http://arxiv.org/abs/2403.13542}
}

@online{maronese_quantum_2021,
  author = {M. Maronese and L. Moro and L. Rocutto and E. Prati},
  title = {Quantum Compiling},
  publisher = {arXiv},
  year = {2021},
  month = {December},
  note = {arXiv:2112.00187 [quant-ph]},
  doi = {10.48550/arXiv.2112.00187},
  url = {http://arxiv.org/abs/2112.00187}
}

@article{wintersperger_neutral_2023,
  author = {K. Wintersperger and F. Dommert and T. Ehmer and A. Hoursanov and J. Klepsch and W. Mauerer and G. Reuber and T. Strohm and M. Yin and S. Luber},
  title = {Neutral Atom Quantum Computing Hardware: Performance and End-User Perspective},
  journal = {EPJ Quantum Technol.},
  volume = {10},
  pages = {32},
  year = {2023},
  month = {December},
  doi = {10.1140/epjqt/s40507-023-00190-1},
  url = {http://arxiv.org/abs/2304.14360}
}

@article{roffe_quantum_2019,
  author = {J. Roffe},
  title = {Quantum Error Correction: An Introductory Guide},
  journal = {Contemp. Phys.},
  volume = {60},
  number = {3},
  pages = {226--245},
  year = {2019},
  month = {July},
  doi = {10.1080/00107514.2019.1667078},
  url = {http://arxiv.org/abs/1907.11157}
}

@online{mugel_use_2020,
  author = {S. Mugel and E. Lizaso and R. Orus},
  title = {Use Cases of Quantum Optimization for Finance},
  publisher = {arXiv},
  year = {2020},
  month = {October},
  note = {arXiv:2010.01312 [q-fin]},
  doi = {10.48550/arXiv.2010.01312},
  url = {http://arxiv.org/abs/2010.01312}
}

@article{orus_practical_2014,
  author = {R. Orus},
  title = {A Practical Introduction to Tensor Networks: Matrix Product States and Projected Entangled Pair States},
  journal = {Ann. Phys.},
  volume = {349},
  pages = {117--158},
  year = {2014},
  month = {October},
  doi = {10.1016/j.aop.2014.06.013},
  url = {http://arxiv.org/abs/1306.2164}
}

@article{sheridan_data-driven_2021,
  author = {E. Sheridan and C. Rhodes and F. Jamet and I. Rungger and C. Weber},
  title = {Data-Driven Dynamical Mean-Field Theory: An Error-Correction Approach to Solve the Quantum Many-Body Problem Using Machine Learning},
  journal = {Phys. Rev. B},
  volume = {104},
  number = {20},
  pages = {205120},
  year = {2021},
  month = {November},
  doi = {10.1103/PhysRevB.104.205120},
  url = {https://link.aps.org/doi/10.1103/PhysRevB.104.205120}
}

@book{kuramoto_quantum_2020,
  author = {Y. Kuramoto},
  title = {Quantum Many-Body Physics: A Perspective on Strong Correlations},
  series = {Lecture Notes in Physics},
  volume = {934},
  publisher = {Springer Japan},
  address = {Tokyo},
  year = {2020},
  doi = {10.1007/978-4-431-55393-9},
  url = {http://link.springer.com/10.1007/978-4-431-55393-9}
}

@article{martens_acceptance_2025,
	title = {Acceptance and {Development} of {Quantum} {Computing} in the {Netherlands} and {Germany}: {Barriers} and {Remedies} {From} a {Multistakeholder} {Perspective}},
	volume = {72},
	issn = {1558-0040},
	shorttitle = {Acceptance and {Development} of {Quantum} {Computing} in the {Netherlands} and {Germany}},
	url = {https://ieeexplore.ieee.org/document/10746616},
	doi = {10.1109/TEM.2024.3493600},
	urldate = {2025-03-10},
	journal = {IEEE Transactions on Engineering Management},
	author = {Martens, Julian and Kumara, Indika and Nucci, Dario Di and Pecorelli, Fabiano and Monsieur, Geert and Tamburri, Damian Andrew and Heuvel, Willem-Jan Van Den},
	year = {2025},
	pages = {62--77},
}

\end{document}